\documentstyle[aps,prl,epsf,preprint,cite,amsmath]{revtex}

\begin{document}
\draft
\title{Neutron Scattering Study of Magnetic Ordering and Excitations in the Doped Spin Gap System Tl(Cu$_{1-x}$Mg$_x$)Cl$_3$}

\author{A. Oosawa,$^1$ M. Fujisawa,$^2$ K. Kakurai,$^1$ and H. Tanaka$^3$}

\address{
$^1$Advanced Science Research Center, Japan Atomic Energy Research Institute,
Tokai, Ibaraki 319-1195, Japan\\
$^2$Department of Physics, Tokyo Institute of Technology, 
Oh-okayama, Meguro-ku, Tokyo 152-8551, Japan \\
$^3$Research Center for Low Temperature Physics, Tokyo Institute of Technology,
Oh-okayama, Meguro-ku, Tokyo 152-8551, Japan \\}

\date{\today}

\maketitle

\begin{abstract}
Neutron elastic and inelastic scattering measurements have been performed in order to investigate the spin structure and the magnetic excitations in the impurity-induced antiferromagnetic ordered phase of the doped spin gap system Tl(Cu$_{1-x}$Mg$_x$)Cl$_3$ with $x=0.03$. The magnetic Bragg reflections indicative of the ordering were observed at ${\pmb Q}=(h, 0, l)$ with integer $h$ and odd $l$ below $T_{\rm N}=3.45$ K. It was found that the spin structure of the impurity-induced antiferromagnetic ordered phase on average in Tl(Cu$_{1-x}$Mg$_x$)Cl$_3$ with $x=0.03$ is the same as that of the field-induced magnetic ordered phase for ${\pmb H} \parallel b$ in the parent compound TlCuCl$_3$. The triplet magnetic excitation was clearly observed in the $a^*$-$c^*$ plane and the dispersion relations of the triplet excitation were determined along four different directions. The lowest triplet excitation corresponding to the spin gap was observed at ${\pmb Q}=(h, 0, l)$ with integer $h$ and odd $l$, as observed in TlCuCl$_3$. It was also found that the spin gap increases steeply below $T_{\rm N}$ upon decreasing temperature. This strongly indicates that the impurity-induced antiferromagnetic ordering coexists with the spin gap state in Tl(Cu$_{1-x}$Mg$_x$)Cl$_3$ with $x=0.03$.
\end{abstract}

\pacs{PACS number 75.10.Jm, 75.25.+z, 75.30.Et, 75.40.Gb, 78.70.Nx}

\section{Introduction}
Recently, the magnetic ordering induced by external field or impurity ions in spin gap systems attracted much interest and thus led to a large number of vivid experimental and theoretical investigations. When a strong magnetic field, which is higher than the gap field $H_{\rm g}$ corresponding to the energy gap $\Delta = g \mu_{\rm B} H_{\rm g}$, is applied in a spin gap system, the energy gap vanishes, and the system can undergo the field-induced magnetic ordering with the help of the three-dimensional interactions. Such field-induced magnetic ordering was observed in many spin gap systems, Cu$_2$(C$_5$H$_{12}$N$_2$)$_2$Cl$_4$ \cite{Hammar1,Hammar2,Chaboussant}, Ni(C$_5$H$_{14}$N$_2$)$_2$N$_3$(ClO$_4$) \cite{Honda1}, Ni(C$_5$H$_{14}$N$_2$)$_2$N$_3$(PF$_6$) \cite{Honda2}, Cu(NO$_3$)$_2$${\cdot}$2.5H$_2$O \cite{Diederix,Bonner2}, (CH$_3$)$_2$CHNH$_3$CuCl$_3$ \cite{Manaka}, (C$_4$H$_{12}$N$_2$)Cu$_2$Cl$_6$ \cite{Stone} and KCuCl$_3$ \cite{OosawaK}. On the other hand, when nonmagnetic ions are substituted for magnetic ions in a spin gap system, the singlet ground state is disturbed so that staggered moments are induced around the impurities. If the induced moments interact through effective exchange interactions, which are mediated by intermediate singlet spins, three-dimensional (3D) long-range order can arise. Such impurity-induced antiferromagnetic ordering was also observed in some spin gap systems, (Cu$_{1-x}$Zn$_x$)GeO$_3$ \cite{Hase}, (Cu$_{1-x}$Mg$_x$)GeO$_3$ \cite{Masuda}, Sr(Cu$_{1-x}$Zn$_x$)$_2$O$_3$ \cite{Azuma} and Pb(Ni$_{1-x}$Mg$_x$)$_2$V$_2$O$_8$ \cite{Uchiyama}. Magnetic ordering due to pinning of lattice dimerization by doped nonmagnetic impurities was also observed in Cu(Ge$_{1-x}$Si$_x$)O$_3$ \cite{Regnault}. In (Cu$_{1-x}$Zn$_x$)GeO$_3$ and Cu(Ge$_{1-x}$Si$_x$)O$_3$ with low $x$, the coexistence of lattice dimerization, associated with the spin gap, and the ordering has been observed in neutron scattering experiments \cite{Martin,Regnault}. The site-impurity effect on spin gap systems has been argued theoretically by many authors \cite{Fukuyama,Miyazaki,Laukamp,Yasuda}, and impurity-induced antiferromagnetic ordering has been demonstrated. Yet there has been no system studied so far, which undergoes both the field and impurity induced magnetic ordering. \par
This paper is concerned with the impurity-induced antiferromagnetic ordering in the Mg-doped spin gap system Tl(Cu$_{1-x}$Mg$_x$)Cl$_3$ where the parent compound TlCuCl$_3$ is known to undergo the field-induced magnetic ordering and has been extensively studied by author's group. The parent compound TlCuCl$_3$ has the monoclinic structure (space group $P2_1/c$) \cite{Takatsu}. The crystal structure is composed of planar dimers of Cu$_2$Cl$_6$, in which Cu$^{2+}$ ions have spin-$\frac{1}{2}$. The dimers are stacked on top of one another to form infinite double chains parallel to the crystallographic $a$-axis. These double chains are located at the corners and center of the unit cell in the $b-c$ plane, and are separated by Tl$^+$ ions. \par
The magnetic ground state of TlCuCl$_3$ is the spin singlet \cite{Takatsu} with the excitation gap $\Delta=7.7$ K \cite{Shiramura,Oosawamag}. From the results of analyses of the dispersion relations obtained by neutron inelastic scattering, it was found that the origin of the spin gap in TlCuCl$_3$ is the strong antiferromagnetic interaction in the chemical dimer Cu$_2$Cl$_6$, and that the neighboring dimers couple via the strong three-dimensional interdimer interactions along the double chain and in the $(1, 0, -2)$ plane, in which the hole orbitals of Cu$^{2+}$ spread \cite{Oosawainela,OosawaASR,Cavadini}. Using the cluster series expansion, the intradimer and the important individual interdimer exchange interactions were obtained \cite{Oosawainela}. The notation of the exchange interactions is given in Fig. \ref{spinstructure}. The main intradimer interaction is denoted as $J$, and then the exchange interaction per bond between spins in dimers separated by a lattice vector $l{\it{\pmb a}} + m{\it{\pmb b}} + n{\it{\pmb c}}$ is denoted as the exchange energy $J_{lmn}$ for pairs of spins at equivalent positions in their respective dimer and as $J_{lmn}'$ for spins at inequivalent positions. \par
The field-induced magnetic ordering, as mentioned above, was observed in TlCuCl$_3$ \cite{Oosawamag,Oosawaheat,Tanakaela}. In the ordering of TlCuCl$_3$, two remarkable features have been found. One of them is that the magnetization has the cusplike minimum at the transition temperature $T_{\rm N}$. Another is that the phase boundary between the paramagnetic phase and the ordered phase can be described by the power law. These features cannot be explained by the mean-field approach from the real space \cite{Tachiki1,Tachiki2}. Nikuni {\it et al.} \cite{Nikuni} suggested that the field-induced magnetic ordering in TlCuCl$_{3}$ can be represented as a Bose-Einstein condensation (BEC) of excited triplets (magnons), and then the above two features were qualitatively well described by the magnon BEC theory based on the Hartree-Fock (HF) approximation. \par
If the magnons undergo BEC at ordering vector ${\pmb Q}_0$ for $H>H_{\rm g}$, then the transverse spin components have long-range order, which is characterized by the same wave vector ${\pmb Q}_0$. The transverse magnetic ordering predicted by the theory \cite{Tachiki1,Tachiki2,Nikuni} was confirmed by the neutron elastic scattering experiments in magnetic fields \cite{Tanakaela}. The obtained spin structure of field-induced ordered phase for ${\pmb H} \parallel b$ in TlCuCl$_3$ is shown in Fig. \ref{spinstructure}. The directions of spins are in the $a-c$ plane which is perpendicular to the applied field. Spins on the same dimers represented by thick lines in Fig. \ref{spinstructure} are antiparallel. Spins are arranged in parallel along a leg in the double chain, and make an angle of $\alpha$ with the $a$-axis. The spins on the same legs in the double chains located at the corner and the center of the unit cell in the $b-c$ plane are antiparallel. This structure is consistent with the sign of the intradimer and the individual interdimer exchange interactions obtained by the neutron inelastic scattering measurements \cite{Oosawainela}. Comparing to the magnetic Bragg intensities, the angle $\alpha$ was estimated as 39$^{\circ}$. \par
More recently, we performed the magnetization measurements in the title compound Tl(Cu$_{1-x}$Mg$_x$)Cl$_3$, and observed the impurity-induced antiferromagnetic ordering, as mentioned above \cite{OosawaMgsus}. It was also confirmed that the easy-axis of the magnetic moments lies in the $(0, 1, 0)$ plane at an angle of $38^{\circ}$ to the $a$-axis. The easy-axis ($\alpha$=38$^{\circ}$) is almost the same as the spin direction ($\alpha$=39$^{\circ}$) in the field-induced ordered phase for the parent compound TlCuCl$_3$. \par
Experimental studies of impurity-induced antiferromagnetic ordering have been performed only on quasi-1D systems to date. Therefore, for comprehensive understanding of impurity-induced antiferromagnetic ordering, it may be necessary to study the ordering on 3D spin gap systems. Also, it is interesting to investigate the relation between the field-induced magnetic ordering and the impurity-induced antiferromagnetic ordering because there is no system so far studied that undergoes both of the orderings except TlCuCl$_3$ to our best knowledge. With these motivations, we performed the neutron elastic and inelastic scattering measurements in Tl(Cu$_{0.97}$Mg$_{0.03}$)Cl$_3$ in order to investigate the spin structure and the magnetic excitations in the impurity-induced antiferromagnetic ordered phase of Tl(Cu$_{1-x}$Mg$_x$)Cl$_3$.

\section{Experimental Details}

The preparation of the single crystal of Tl(Cu$_{1-x}$Mg$_x$)Cl$_3$ has been reported in reference \cite{OosawaMgsus}. Neutron elastic and inelastic scattering measurements were performed using the JAERI-TAS1 and JAERI-LTAS, which are thermal and cold neutron spectrometers respectively, installed at JRR-3M, in Tokai. For the elastic scattering, the constant-${\pmb k}_i$ mode was taken with a fixed incident neutron energy $E_i$ of 14.7 meV and collimations were set as open-40'-80'-80' on JAERI-TAS1. For the inelastic scattering, the constant-${\pmb k}_f$ mode was taken with a fixed final neutron energy $E_f$ of 14.7 meV and collimations were set as open-80'-80'-80' on JAERI-TAS1. In order to investigate the low energy excitation with the higher energy resolution, the inelastic scattering, in which the constant-${\pmb k}_f$ mode was taken with a fixed final neutron energy $E_f$ of 5.0 meV and collimations were set as 24'-80'-80'-80', were also carried out on JAERI-LTAS. A pyrolytic graphite filter was placed to suppress the higher order contaminations. We used a sample with a volume of approximately 1 cm$^3$. The sample was mounted in an ILL-type orange cryostat with its $a^*$- and $c^*$-axes in the scattering plane. The crystallographic parameters were determined as $a^*=1.6139$ 1/$\rm{\AA}$, $c^*=0.71554$ 1/$\rm{\AA}$ and $\cos\beta^*=0.0956$ at helium temperatures. 

\section{Results and Discussions}

\subsection{Spin Structure}

\begin{figure}[ht]
\vspace*{5cm}
\epsfxsize=120mm
\centerline{\epsfbox{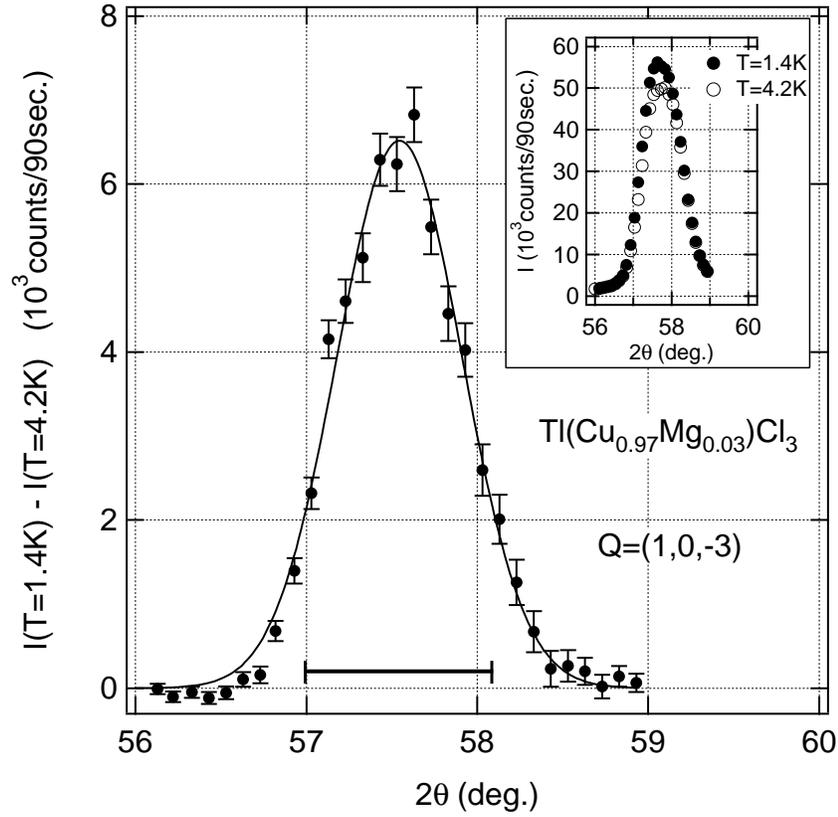}}
\vspace*{1cm}
	\caption{The intensity difference $I(T=1.4$K$)-I(T=4.2$K$)$ of the ${\theta}-2{\theta}$ scans for ${\pmb Q}=(1, 0, -3)$ reflection in Tl(Cu$_{0.97}$Mg$_{0.03}$)Cl$_3$. The inset shows the ${\theta}-2{\theta}$ scans for ${\pmb Q}=(1, 0, -3)$ reflection measured at $T=1.4$ and 4.2 K in Tl(Cu$_{0.97}$Mg$_{0.03}$)Cl$_3$. The horizontal bar indicates the calculated instrumental resolution widths.}
	\label{Q10-3profile}
\end{figure}

\begin{figure}[ht]
\vspace*{5cm}
\epsfxsize=120mm
\centerline{\epsfbox{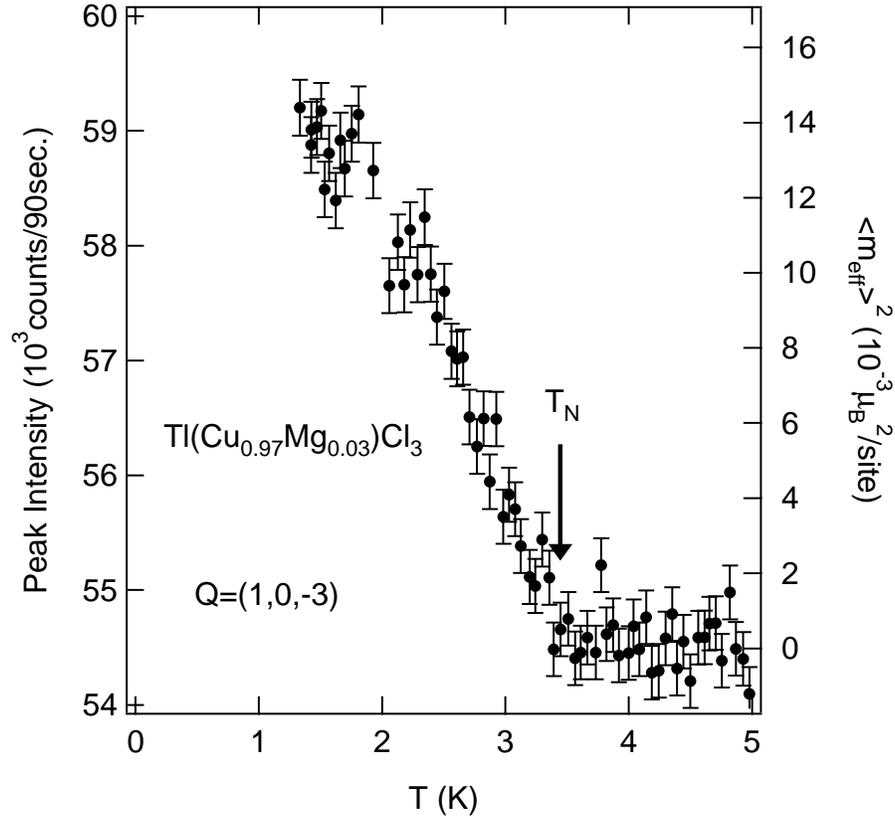}}
\vspace*{1cm}
	\caption{Temperature dependence of the magnetic Bragg peak intensity for ${\pmb Q}=(1, 0, -3)$ reflection in Tl(Cu$_{0.97}$Mg$_{0.03}$)Cl$_3$. The square of the effective magnetic moment averaged per site $\langle m_{\rm eff} \rangle^2$ is shown on the right abscissa.}
	\label{Q10-3temdep}
\end{figure}

The inset of Fig. \ref{Q10-3profile} shows the ${\theta}-2{\theta}$ scans for ${\pmb Q}=(1, 0, -3)$ reflection measured at $T=1.4$ and 4.2 K in Tl(Cu$_{0.97}$Mg$_{0.03}$)Cl$_3$. Figure \ref{Q10-3profile} depicts the intensity difference $I(T=1.4$K$)-I(T=4.2$K$)$ of these scans. A resolution limited additional intensity at low temperature can be clearly seen. Figure \ref{Q10-3temdep} shows the temperature dependence of the peak intensity at ${\pmb Q}=(1, 0, -3)$. The rapid increase of the intensity below $T_{\rm N}=3.45$ K is observed. This transition temperature is in agreement with the phase diagram of this doped sample obtained from the magnetization measurements \cite{OosawaMgsus}. We thus infer that the additional intensity at low $T$ is of magnetic origin. These magnetic intensities at low temperature are observed at $Q=(h, 0, l)$ for integer $h$ and odd $l$ in the doped sample. These reciprocal points are the same as those, at which the magnetic Bragg peaks indicative of the field-induced transverse N\'{e}el ordering were observed in TlCuCl$_3$. In order to determine the spin structure of the impurity-induced antiferromagnetic ordered phase in Tl(Cu$_{0.97}$Mg$_{0.03}$)Cl$_3$, we measured the integrated intensities of nine Bragg reflections at $T=1.4$ and 4.2 K. The integrated intensities of the magnetic Bragg reflections were obtained by fitting the difference of the observed intensities at $T=1.4$ and 4.2 K using a Gaussian function, as shown in Fig. \ref{Q10-3profile}. The results are summalized in Table \ref{table2}. The magnetic Bragg reflections are resolution limited ({\it e.g.} see Fig. \ref{Q10-3profile}) and correspond to a correlation length greater than 150 $\AA$. This indicates that the ordering is of a long range nature. The intensities of nuclear Bragg reflections were also measured at ${\pmb Q}=(h, 0, l)$ with even $l$ in order to estimate the magnitude of the effective magnetic moment averaged per site. Because the parent compound TlCuCl$_3$ belongs to the space group $P2_1/c$, nuclear peaks are expected only at ${\pmb Q}=(h, 0, l)$ with even $l$. However, as shown in the inset of Fig. \ref{Q10-3profile}, the nuclear peaks, though weak, were observed also for odd $l$ at $T=4.2$ K. In the pure TlCuCl$_3$, very weak nuclear peaks have been observed for odd $l$ \cite{Tanakaela} which may be attributed to the impurity induced small distortion lowering the local symmetry. The intensities of the odd $l$ nuclear peaks observed in the doped compound Tl(Cu$_{0.97}$Mg$_{0.03}$)Cl$_3$ was larger than those observed in the pure TlCuCl$_3$, but still smaller than the allowed even $l$ peaks in Tl(Cu$_{0.97}$Mg$_{0.03}$)Cl$_3$. We checked that the allowed nuclear Bragg intensities in the present measurements are consistent with the nuclear structure factor evaluated from the crystal structure of TlCuCl$_3$ for the weak reflection, in which the extinction effect is almost ineffective. As shown in reference \cite{OosawaMgsus}, it is also confirmed that the magnetic susceptibilities of this doped sample behave the same as TlCuCl$_3$ above $T=7$ K, indicating the absence of any structural phase transition. We therefore infer that the nuclear peaks observed at odd $l$ are due to the local disturbance of lattice caused by the Mg doping. One possible reason for such a local disturbance could be caused, for instance, by the difference between the Jahn-Teller (JT) distorted CuCl$_6$ octahedra and non JT active ion containing MgCl$_6$ octahedra, so that the extinction law for odd $l$ could be locally broken. Although more detailed informations of the observed reflections at odd $l$ can be obtained by powder diffraction measurements at low temperature, we believe that the magnetic results presented here will be not affected. To refine the magnetic structure, we therefore used the atomic coordinates of TlCuCl$_3$ \cite{Tanakaela} and the nuclear scattering lengths $b_{\rm Tl}=0.878$, $b_{\rm Cu}=0.772$ and $b_{\rm Cl}=0.958$ in the unit of 10$^{-12}$ cm \cite{Sears}. The magnetic form factors of Cu$^{2+}$ were taken from reference \cite{Brown}. The extinction effect was evaluated by comparing observed and calculated intensities for various nuclear Bragg reflections. \par
Because the magnetic Bragg reflections were observed at ${\pmb Q}=(h, 0, l)$ with integer $h$ and odd $l$, the antiferromagnetic ordering as shown in Fig. \ref{spinstructure} is expected. As mentioned in Introduction, it is suggested that the spin lies in the $a-c$ plane from the magnetization measurements \cite{OosawaMgsus}. Hence the free parameters in this calculation are only $\alpha$ and $C$, which are the angle between the spins and the $a$-axis, and the scale factor, respectively. The calculated intensities of the best fit with $\alpha = 34.0^{\circ} \pm 4.7^{\circ}$ are indicated in Table \ref{table2} and the overall agreement can be considered as being satisfactory. The obtained angle $\alpha$ is consistent with the magnetization measurements of Tl(Cu$_{1-x}$Mg$_x$)Cl$_3$, and is, with in error bars, the same as that in the spin structure of the field-induced ordered phase of TlCuCl$_3$ for ${\pmb H} \parallel b$. Hence we conclude that the spin structure of the impurity-induced antiferromagnetic ordered phase on average in Tl(Cu$_{0.97}$Mg$_{0.03}$)Cl$_3$ is the same as that of the field-induced magnetic ordered phase for ${\pmb H} \parallel b$ in the parent compound TlCuCl$_3$ ("on average" means that the magnitude of the ordered moments in the impurity-induced antiferromagnetic ordered phase may be not uniform.). \par
Comparing magnetic peak intensities with those of nuclear reflections, the magnitude of the effective magnetic moment averaged per site at $T=1.4$ K was evaluated as $\langle m_{\rm eff} \rangle =g{\mu}_{\rm B}\langle S_{\rm eff} \rangle = 0.12(1)$ ${\mu}_{\rm B}$. On the right abscissa of Fig. \ref{Q10-3temdep}, we show $\langle m_{\rm eff} \rangle^2$. This value is much smaller than 1 $\mu_{\rm B}$, which is the full moment of Cu$^{2+}$, and close to that in the impurity-induced ordered phase observed in the other doped-Cu$^{2+}$ compounds, (Cu$_{1-x}$Zn$_x$)GeO$_3$ \cite{Martin,Haseneu} and (Cu$_{1-x}$Mg$_x$)GeO$_3$ \cite{Nakao}. As shown in Fig. \ref{Q10-3temdep}, no pronounced diffuse scattering can be seen, and the increase of the magnetic peak intensity is almost linear in the vicinity of $T_{\rm N}$.

\subsection{Magnetic Excitations}

\begin{figure}[ht]
\vspace*{5cm}
\begin{minipage}{7.5cm}
 \epsfxsize=75mm
  \centerline{\epsfbox{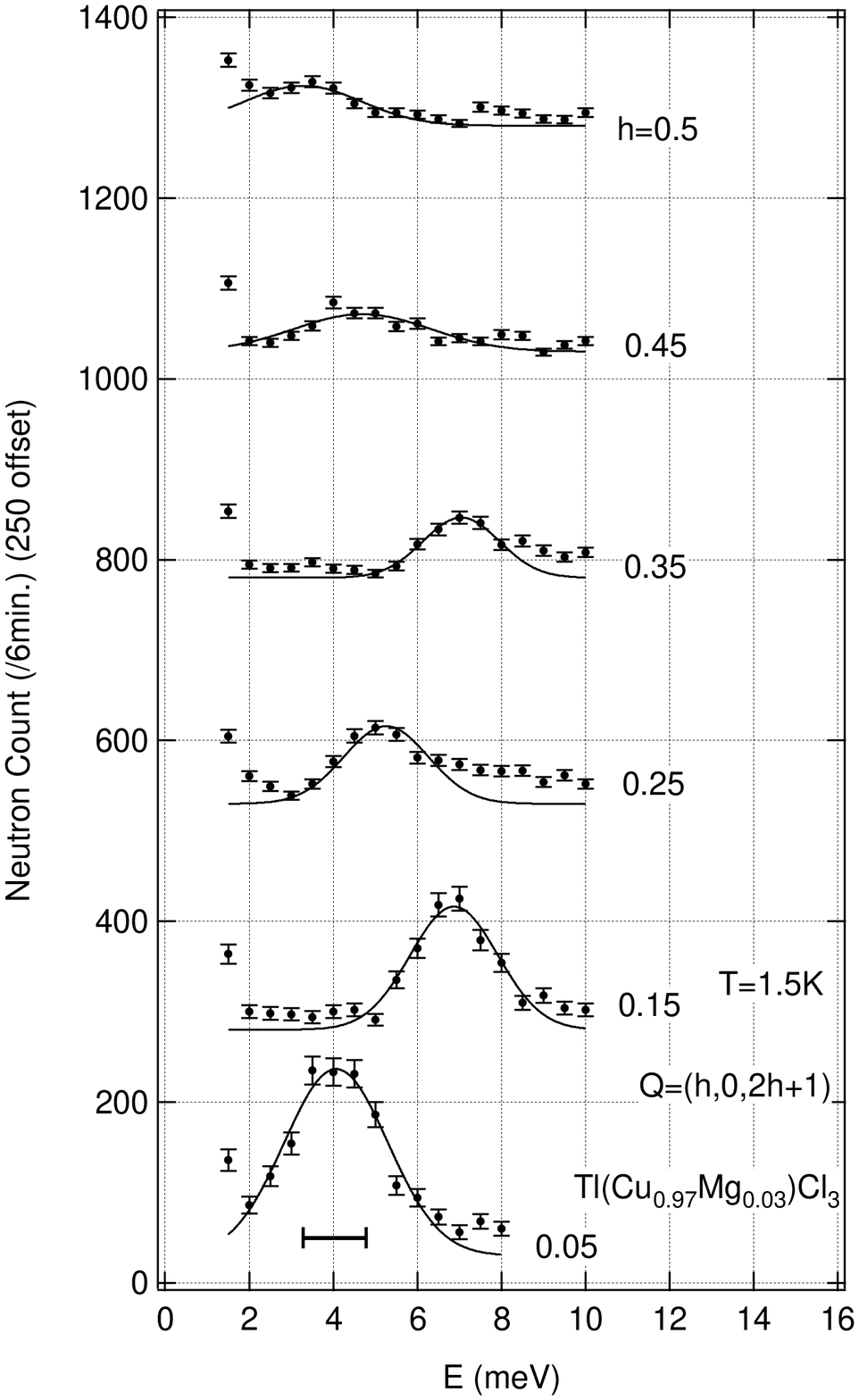}}
\begin{center}
(a)
\end{center}
\end{minipage}
\begin{minipage}{7.5cm}
 \epsfxsize=75mm
  \centerline{\epsfbox{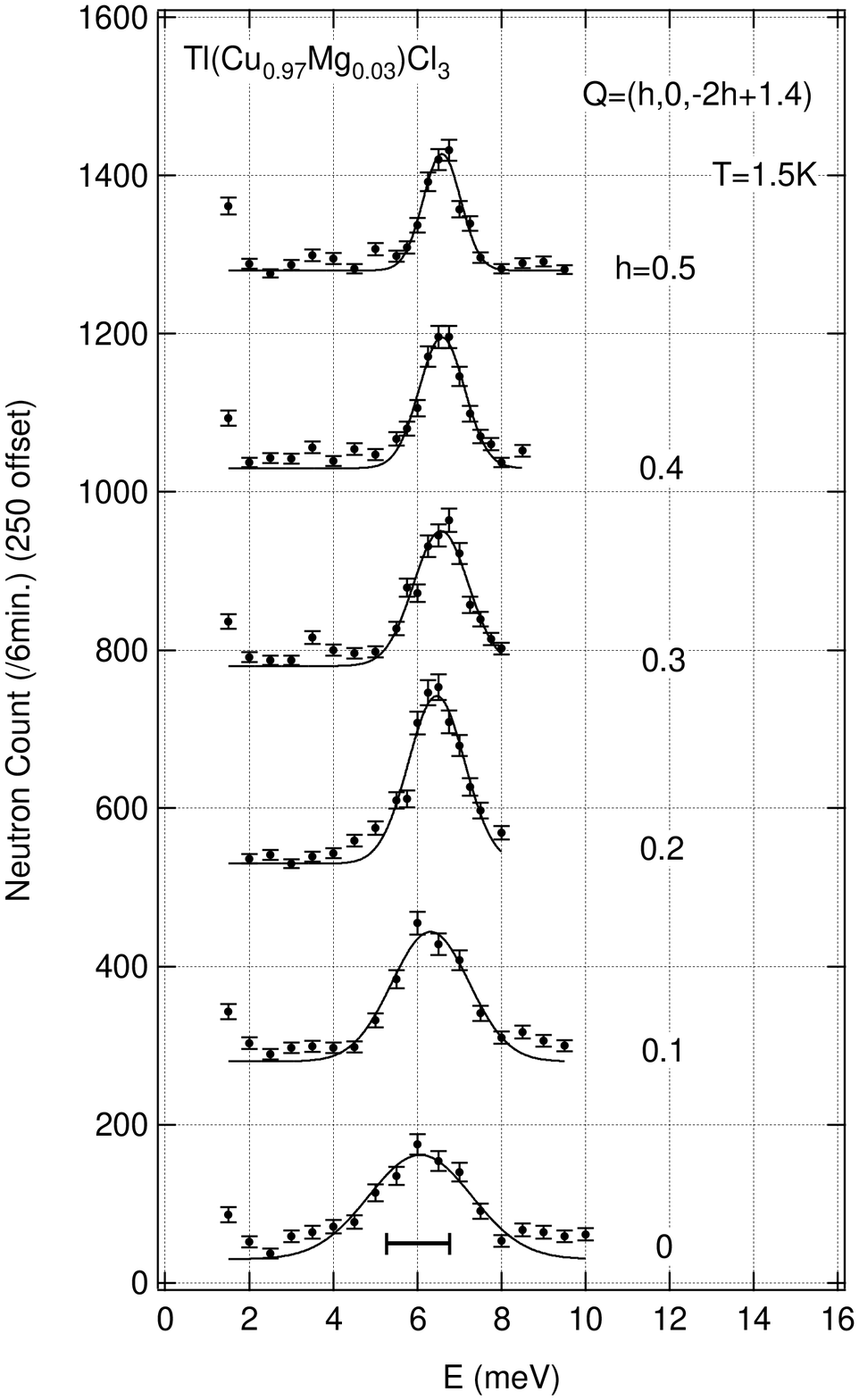}}
\begin{center}
(b)
\end{center}
\end{minipage}
\vspace{1cm}
    \caption{Profiles of the constant-${\pmb Q}$ energy scans in Tl(Cu$_{0.97}$Mg$_{0.03}$)Cl$_3$ for $\it{\pmb Q}$ along (a) $(h, 0, 2h+1)$ and (b) $(h, 0, -2h+1.4)$ with $0\leq h\leq 0.5$. The solid lines are fits using a Gaussian function. The horizontal bars indicate the calculated instrumental resolution widths.} 
    \label{inelacount}
\end{figure}

\begin{figure}[ht]
\vspace*{5cm}
\epsfxsize=120mm
\centerline{\epsfbox{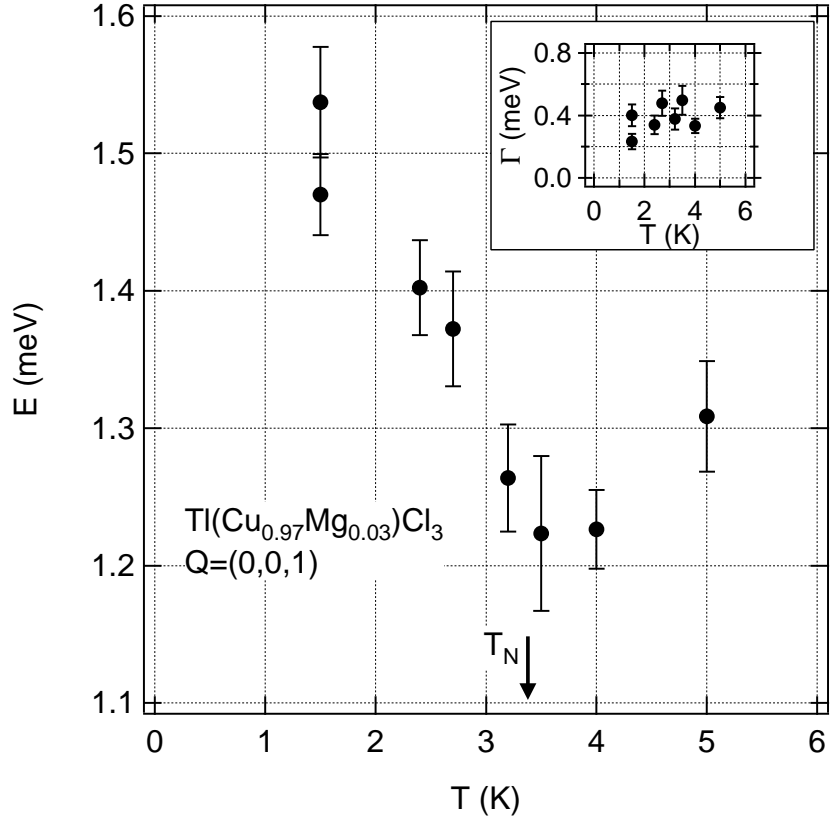}}
\vspace*{1cm}
	\caption{Temperature dependence of the energy of the magnetic excitation at ${\pmb Q}=(0, 0, 1)$. The inset shows the temperature dependence of the Lorentzian width $\Gamma$.}
	\label{001temdep}
\end{figure}

Constant-${\pmb Q}$ energy scan profiles were measured along $(h, 0, 1)$, $(0, 0, l)$, $(h, 0, 2h+1)$ and $(h, 0, -2h+1.4)$ with $0 \le h \le 0.5$ and $1 \le l \le 2$ at $T=1.5$ K, which is lower than the N\'{e}el temperature $T_{\rm N}=3.45$ K in Tl(Cu$_{0.97}$Mg$_{0.03}$)Cl$_3$. Figure \ref{inelacount} shows the representative scan profiles for ${\it{\pmb Q}}=(h, 0, 2h+1)$ and $(h, 0, -2h+1.4)$, in which the magnetic excitation is most dispersive and less dispersive, respectively. A well-defined single excitation can be observed in almost all scans. The scan profiles were fitted with a Gaussian function to evaluate the excitation energy, as shown by the solid lines in Fig. \ref{inelacount}. The obtained scan profiles in Tl(Cu$_{0.97}$Mg$_{0.03}$)Cl$_3$ were almost the same as those obtained in the parent compound TlCuCl$_3$ \cite{Oosawainela,OosawaASR,Cavadini}, and we confirmed that the origin of the observed excitation is magnetic from the temperature dependence. Hence we can expect that the observed excitation is the triplet magnetic excitation in the impurity-induced ordered state. The dispersion relation $\omega({\pmb Q})$ of the excitation obtained for the $a^*-c^*$ scattering plane at $T=1.5$ K is summarized in Fig. \ref{ineladis}. The dispersion curves of the triplet excitation in the parent compound TlCuCl$_3$ \cite{Oosawainela,OosawaASR,Cavadini} is also depicted in Fig. \ref{ineladis} by the dashed lines. As shown in Fig. \ref{ineladis}, there is not much difference between the dispersion relations of Tl(Cu$_{0.97}$Mg$_{0.03}$)Cl$_3$ and those of TlCuCl$_3$, except in the vicinity of ${\pmb Q}=(0, 0, 1)$. Based on the present results, it is evident that the periodicity of the triplet excitation in Tl(Cu$_{0.97}$Mg$_{0.03}$)Cl$_3$ is the same as that of the nuclear reciprocal lattice along the $a^*$-axis, but doubled along the $c^*$-axis, as observed in TlCuCl$_3$ \cite{Oosawainela,OosawaASR,Cavadini}. Hence, it is deduced that the lowest triplet excitation, corresponding to the spin gap, occurs at ${\it{\pmb Q}}=(h, 0, l)$ with integer $h$ and odd $l$ in the $a^*-c^*$ plane. Figure \ref{001count} shows the constant-${\pmb Q}$ energy scan profiles in Tl(Cu$_{0.97}$Mg$_{0.03}$)Cl$_3$ for ${\pmb Q}=(0, 0, 1)$ at various temperatures. A well-defined single excitation was also observed at all temperatures. Hence it is confirmed that the spin gap remains even in the impurity-induced ordered phase. Because the excitation was also observed at $T=5.2$ K, which is higher than $T_{\rm N}$, the excitation is not spin-wave excitation by the antiferromagnetic ordering, but the triplet excitation from the ground state. The horizontal bars in Fig. \ref{001count} denote the calculated instrumental resolution widths. The broadening of the spectra can be seen. Far from ${\pmb Q}=(0, 0, 1)$, it was confirmed that almost all triplet excitations have widths equal to the resolution limit, though studied with coarser resolution. Such behaviors were also reported in (Cu$_{1-x}$Zn$_x$)GeO$_3$ \cite{Martin} and Cu(Ge$_{1-x}$Si$_x$)O$_3$ \cite{Hirota}. The solid lines in Fig. \ref{001count} are fits to the dispersion relations convoluted with the instrumental resolution. A Lorentzian function for the magnetic excitation is used. The free parameters are $I$, $\Gamma$ and $E_{(0,0,1)}$, which are the integrated intensity, the Lorentzian width and the spin gap, respectively. The shape of the dispersion relation in the vicinity of ${\pmb Q}$=(0, 0, 1) was fixed to that obtained from the RPA approximation (see below). The temperature dependence of the energy of the triplet excitation at ${\pmb Q}=(0, 0, 1)$ is shown in Fig. \ref{001temdep}. The rapid increase of the excitation energy can be clearly seen below $T_{\rm N}=3.45$ K. The triplet excitation depends on temperature due to the temperature dependence of the occupation difference between the singlet and triplet states \cite{Cavadinitemp}. However, the increase of temperature causes to flatten the dispersion relation due to the suppression of the interdimer correlations by the decrease of the occupation difference, namely the energy of the lowest excitation corresponding to the spin gap should rather decrease upon decreasing temperature. Hence, we can expect that the rapid increase of the spin gap is caused by the impurity-induced antiferromagnetic ordering and can conclude that the impurity-induced antiferromagnetic ordering coexists with the spin gap in Tl(Cu$_{0.97}$Mg$_{0.03}$)Cl$_3$. The energy of the lowest excitation is $E=1.3$ meV at $T=5.2$ K, which is higher than $T_{\rm N}$, and is a little larger than that of 0.65 $\sim$ 0.8 meV estimated for the pure TlCuCl$_3$ \cite{Oosawainela,OosawaASR,Cavadini}. This indicates that the spin gap is also enhanced by the Mg doping, irrespective of the ordering. The temperature dependence of $\Gamma$ is also shown in inset of Fig. \ref{001temdep}. No significant variation of $\Gamma$ can be seen. \par
We fit the obtained dispersion relations, as shown in Fig. \ref{ineladis}, by the following equation obtained from the RPA approximation, which is applied to the parent compound TlCuCl$_3$ \cite{Oosawainela,OosawaASR,Cavadini}
\begin{equation}
\label{RPA}
\omega_{\pm}({\it{\pmb Q}})=\sqrt{J^2+ 2J \delta \omega_{\pm}({\it{\pmb Q}})}\ ,
\end{equation}
where
\begin{eqnarray}
\label{omega}
\delta \omega_{\pm}({\it{\pmb Q}}) &=&  
    [J_{(100)}^{\rm eff}{\cos}(2{\pi}h) +
     J_{(200)}^{\rm eff}{\cos}(4{\pi}h) \nonumber \\
     & & {} + J_{(201)}^{\rm eff} {\cos}\{ 2{\pi}(2h+l)\}] \nonumber \\
     & & {} {\pm} 2 [J_{(1\frac{1}{2}\frac{1}{2})}^{\rm eff} {\cos}\{{\pi}(2h+l)\}{\cos}({\pi}k) \nonumber \\
     & & {} + 
 J_{(0\frac{1}{2}\frac{1}{2})}^{\rm eff} {\cos}({\pi}k){\cos}({\pi}l)] \ .
\end{eqnarray}
In the equations, $J$ denotes the main intradimer exchange interaction, and $J_{(lmn)}^{\rm eff}$ denote the effective interactions between dimers separated by a lattice vector $l{\it{\pmb a}} + m{\it{\pmb b}} + n{\it{\pmb c}}$ and are expressed as
\begin{eqnarray}
J^{\rm eff}_{(100)} &= & \frac{1}{2}\left(2J_{(100)}-J'_{(100)}\right)\ , \nonumber \\
J^{\rm eff}_{(200)} &= & \frac{1}{2}\left(2J_{(200)}-J'_{(200)}\right)\ , \nonumber \\ 
J^{\rm eff}_{(1\frac{1}{2}\frac{1}{2})} &= & \frac{1}{2}\left(J_{(1\frac{1}{2}\frac{1}{2})}-J'_{(1\frac{1}{2}\frac{1}{2})}\right)\ , \nonumber \\
J^{\rm eff}_{(0\frac{1}{2}\frac{1}{2})} &= & \frac{1}{2}\left(J_{(0\frac{1}{2}\frac{1}{2})}-J'_{(0\frac{1}{2}\frac{1}{2})}\right)\ ,\nonumber \\
J^{\rm eff}_{(201)} &= & -\frac{1}{2}J'_{(201)}\ ,
\end{eqnarray}
where $J_{lmn}$ and $J_{lmn}'$ are shown in Fig. \ref{spinstructure}. Under the present experimental condition, {\it i.e.}, $\it{\pmb Q}$ in the $a^*-c^*$ plane, the $\omega_{+}({\it{\pmb Q}})$ branch gives the only nonvanishing contribution. The solid lines in Fig. \ref{ineladis} indicate the calculated results with the exchange parameters, as shown in Table \ref{table3}. The experimental dispersion curves can be reproduced well by the fitting. Compared to the parent compound TlCuCl$_3$, the magnitude of the intradimer interaction $J$ becomes slightly larger, while that of the important effective interdimer interactions $J_{(100)}^{\rm eff}$, $J_{(1\frac{1}{2}\frac{1}{2})}^{\rm eff}$ and $J_{(201)}^{\rm eff}$ become slightly smaller. \par
Because the impurity-induced antiferromagnetic ordering occurs, we can expect that the antiferromagnetic spin-wave excitation is observed together with the triplet excitation, as observed in (Cu$_{1-x}$Zn$_x$)GeO$_3$ \cite{Martin} and Cu(Ge$_{1-x}$Si$_x$)O$_3$ \cite{Hirota}. However, no antiferromagnetic spin-wave excitation emerging from the antiferromagnetic zone center could be observed in the present neutron scattering experiments though the antiferromagnetic resonance has been observed in this system by the ESR measurements \cite{Shindo}. In both (Cu$_{1-x}$Zn$_x$)GeO$_3$ \cite{Martin} and Cu(Ge$_{1-x}$Si$_x$)O$_3$ \cite{Hirota}, it was reported that the energy of the antiferromagnetic spin-wave excitation is very small compared with that of the triplet excitation and that the line width broadens rapidly with increasing $q$, which is distance from the antiferromagnetic zone center in reciprocal space. Hence it may be that we could not observe the low energy spin-wave excitation due to the large incoherent scattering by the chlorine near the zone center ${\pmb Q}=(0, 0, 1)$ and due to the rapid broadening of the spectra far from ${\pmb Q}=(0, 0, 1)$. \par
The present measurements were performed in zero magnetic field. We plan to perform the neutron scattering measurements for Tl(Cu$_{1-x}$Mg$_x$)Cl$_3$ in a magnetic field. Especially, when a magnetic field higher than the gap field $H_{\rm g}$ is applied at the temperature lower than $T_{\rm N}$, the spin gap between the ground state, in which the impurity-induced antiferromagnetic ordering occurs, and one of the triplet excited states vanishes due to the Zeeman interaction, and then an additional phase transition may occur, similar to the field-induced magnetic ordering in the parent compound TlCuCl$_3$. \par 

\section{Conclusion}

\begin{figure}[ht]
\vspace*{5cm}
\epsfxsize=120mm
\centerline{\epsfbox{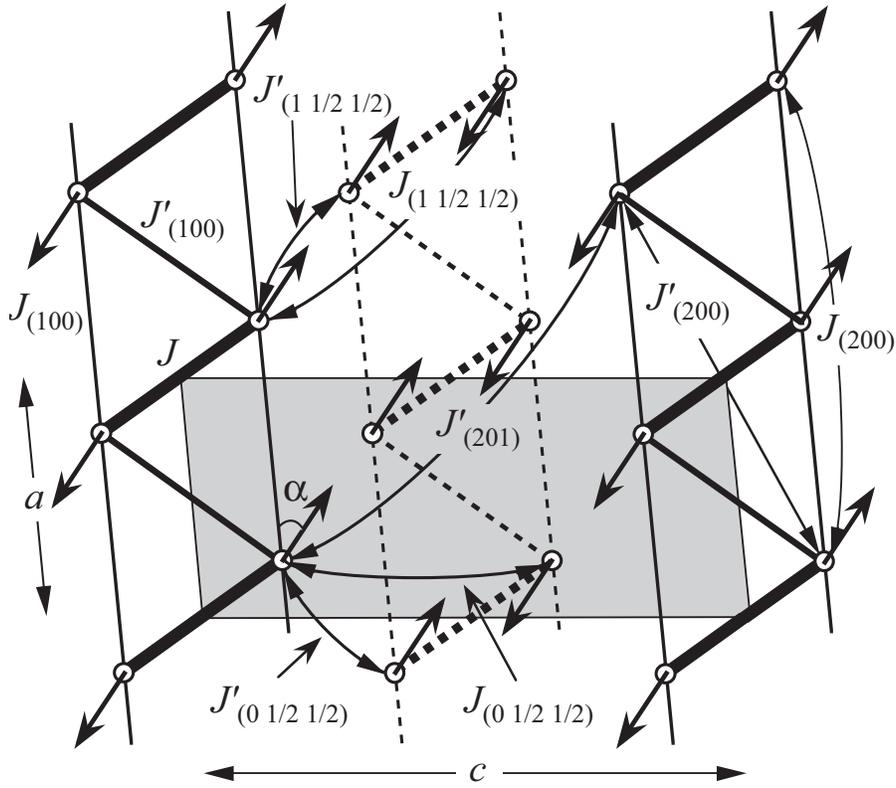}}
\vspace*{1cm}
	\caption{The exchange interactions and the spin structure observed in the field-induced ordered phase of TlCuCl$_3$ for $H \parallel b$ and the impurity-induced ordered phase of Tl(Cu$_{0.97}$Mg$_{0.03}$)Cl$_3$. The double chain located at the corner and the center of the chemical unit cell in the $b-c$ plane are represented by solid and dashed lines, respectively. The shaded area is the chemical unit cell in the $a-c$ plane. The angle $\alpha$ between the spins and the $a$-axis in each phases is expressed in the text.}
	\label{spinstructure}
\end{figure}

\begin{figure}[ht]
\vspace*{3cm}
\begin{minipage}{7.5cm}
\begin{center}
 \epsfxsize=70mm
  \centerline{\epsfbox{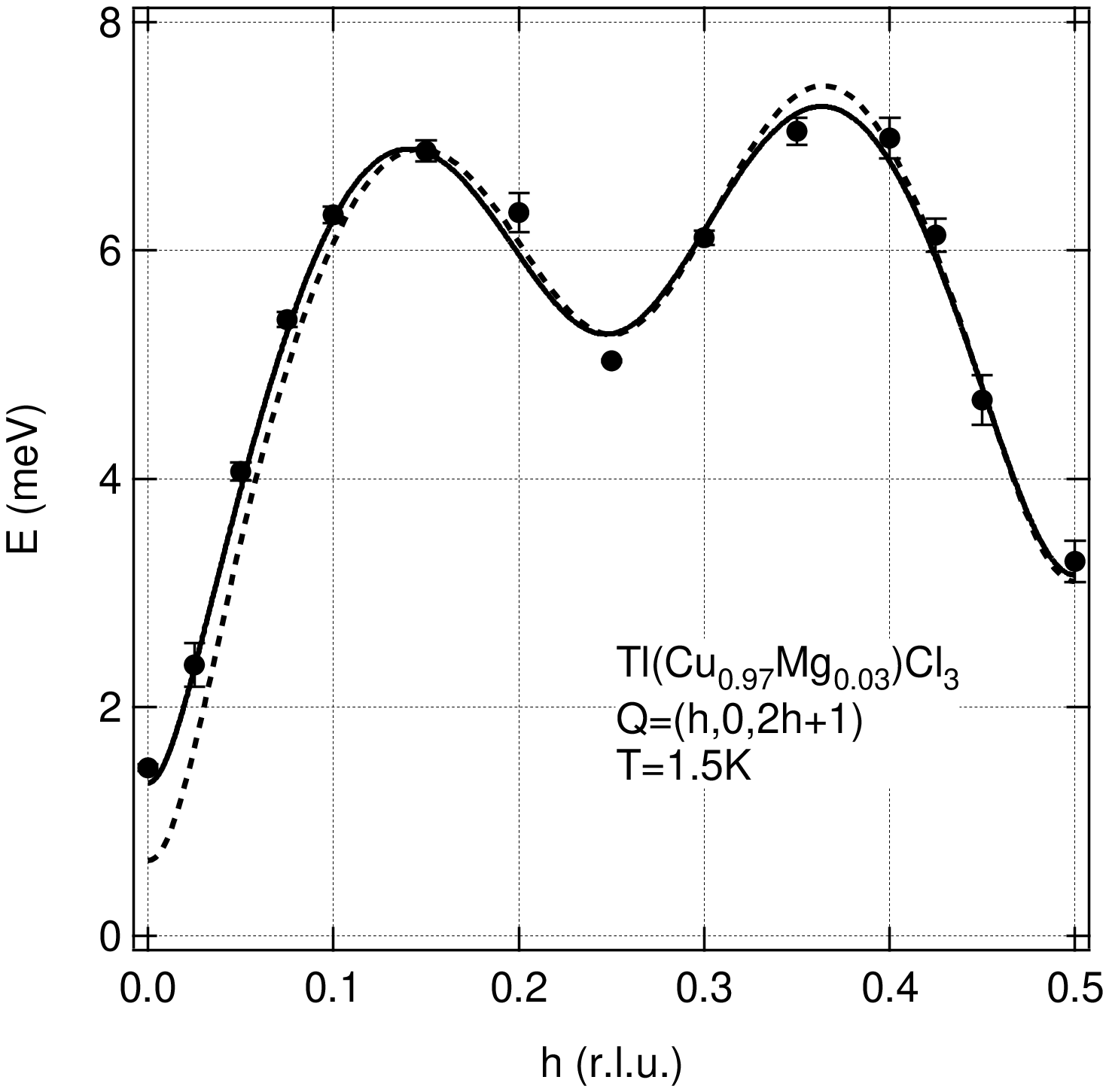}}
(a)
\end{center}
\end{minipage}
\begin{minipage}{7.5cm}
\begin{center}
 \epsfxsize=70mm
  \centerline{\epsfbox{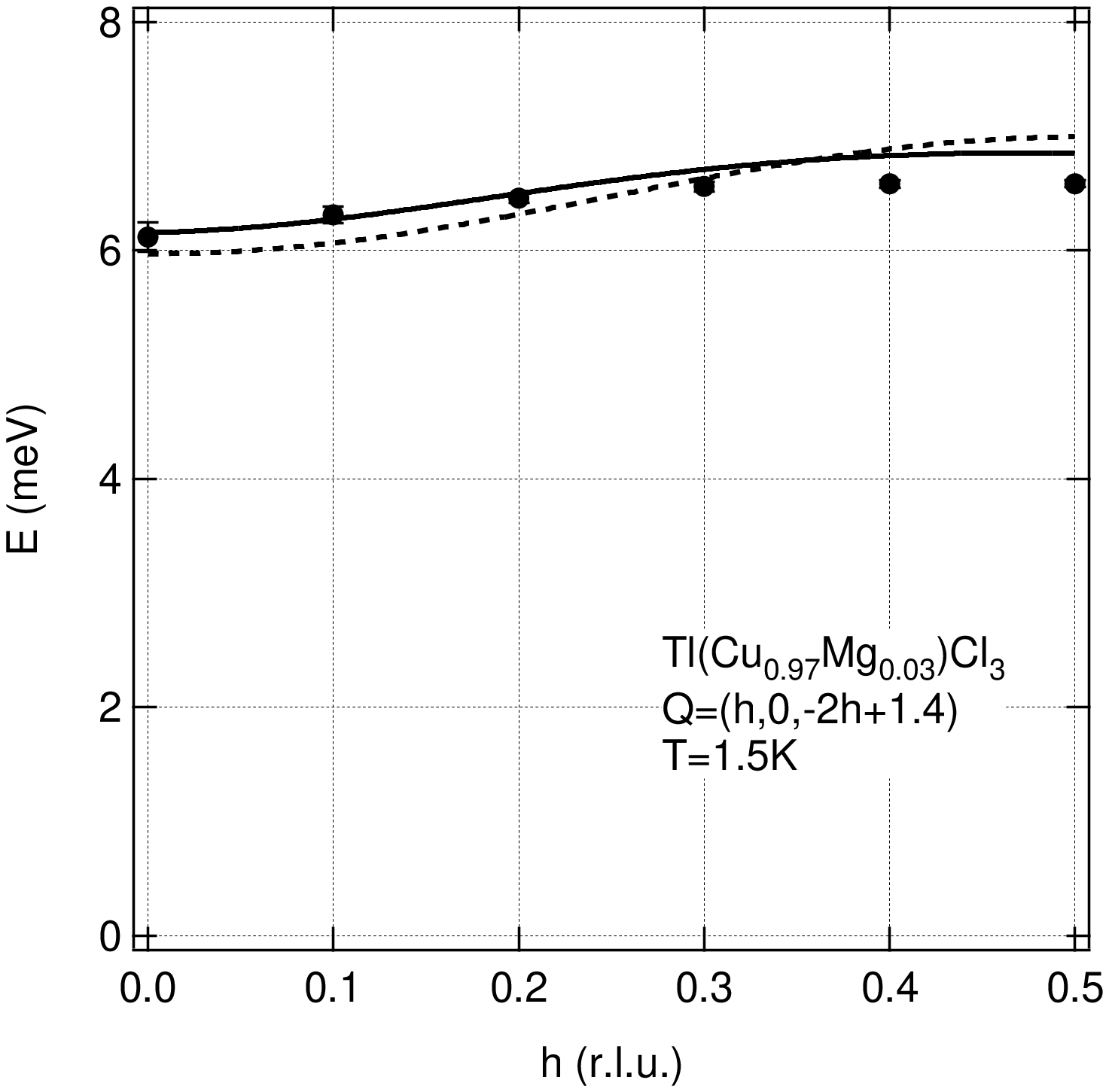}}
(b)
\end{center}
\end{minipage}\par
\begin{minipage}{7.5cm}
\begin{center}
 \epsfxsize=70mm
  \centerline{\epsfbox{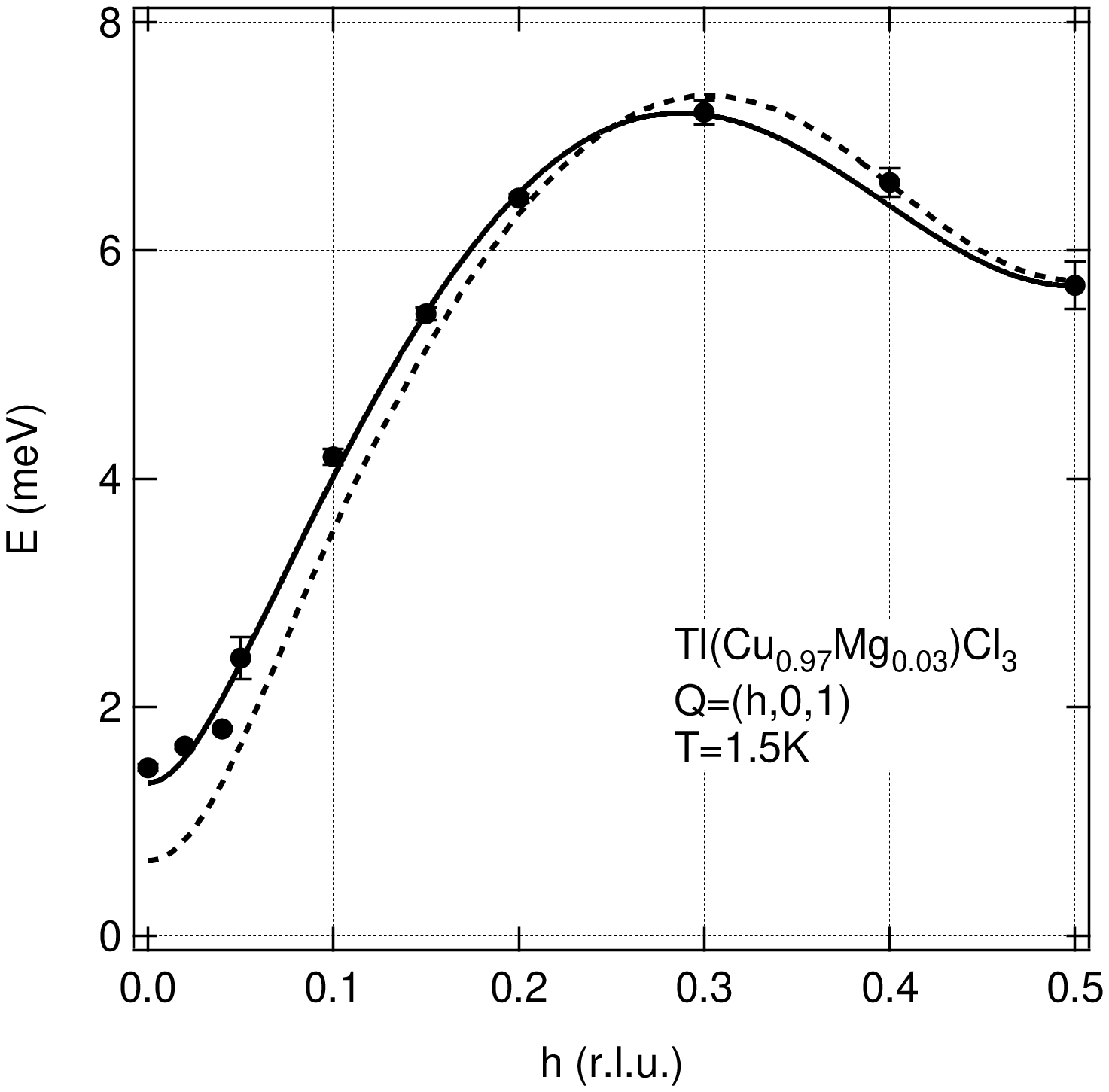}}
(c)
\end{center}
\end{minipage}
\begin{minipage}{7.5cm}
\begin{center}
 \epsfxsize=70mm
  \centerline{\epsfbox{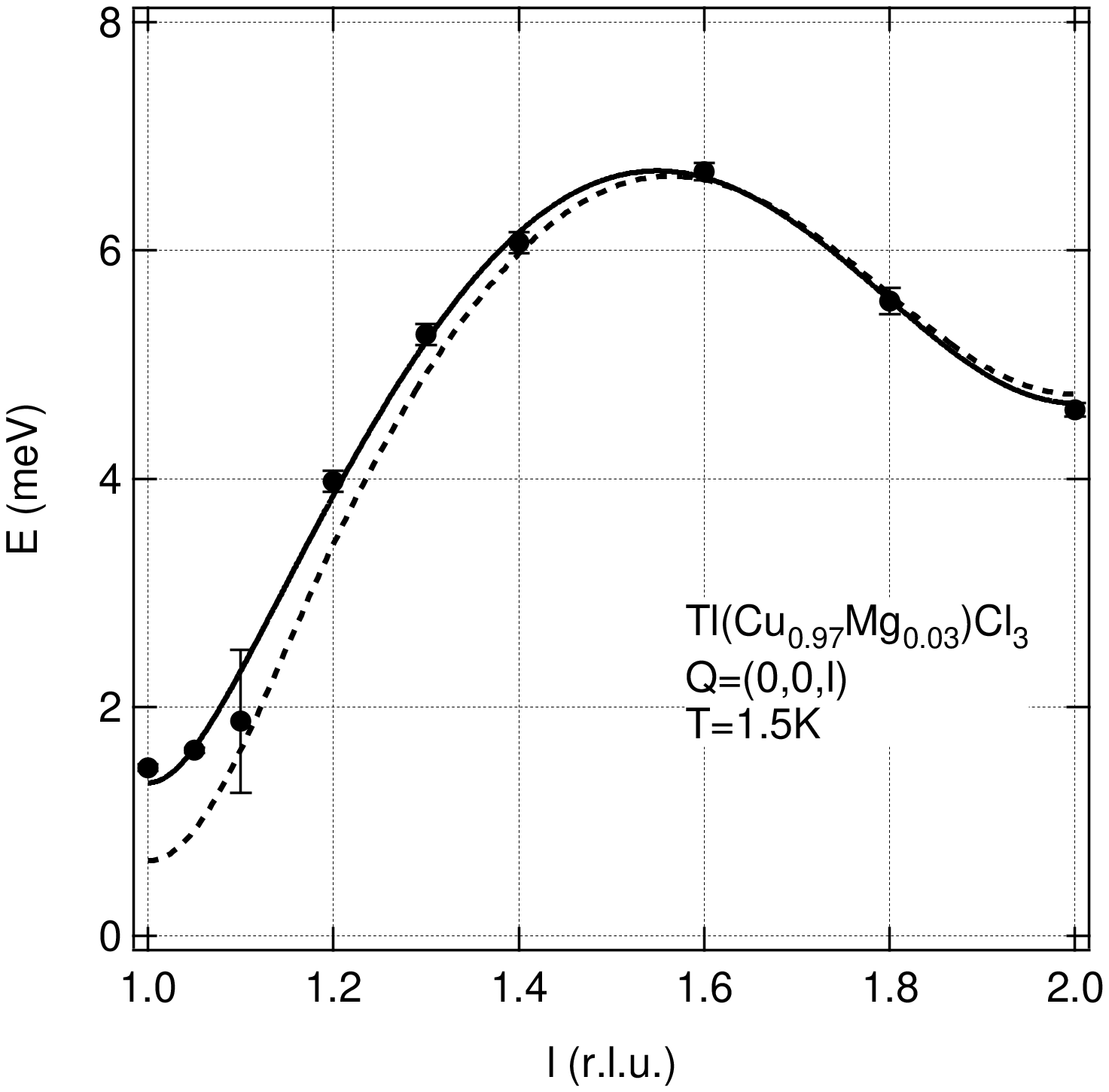}}
(d)
\end{center}
\end{minipage}
\vspace{1cm}
    \caption{Dispersion relations $\omega(\it{\pmb Q})$ in Tl(Cu$_{0.97}$Mg$_{0.03}$)Cl$_3$ for $\it{\pmb Q}$ along (a) $(h, 0, 2h+1)$, (b) $(h, 0, -2h+1.4)$, (c) $(h, 0, 1)$ and (d) $(0, 0, l)$ with $0\leq h\leq 0.5$ and $1\leq l\leq 2$ at $T=1.5$ K. Solid lines are the dispersion curves calculated by RPA approximation using eq. (\ref{RPA}). Dashed lines are the dispersion curves of the triplet excitation in the parent compound TlCuCl$_3$ \cite{Oosawainela}.} 
    \label{ineladis}
\end{figure}

\begin{figure}[ht]
\epsfxsize=120mm
\centerline{\epsfbox{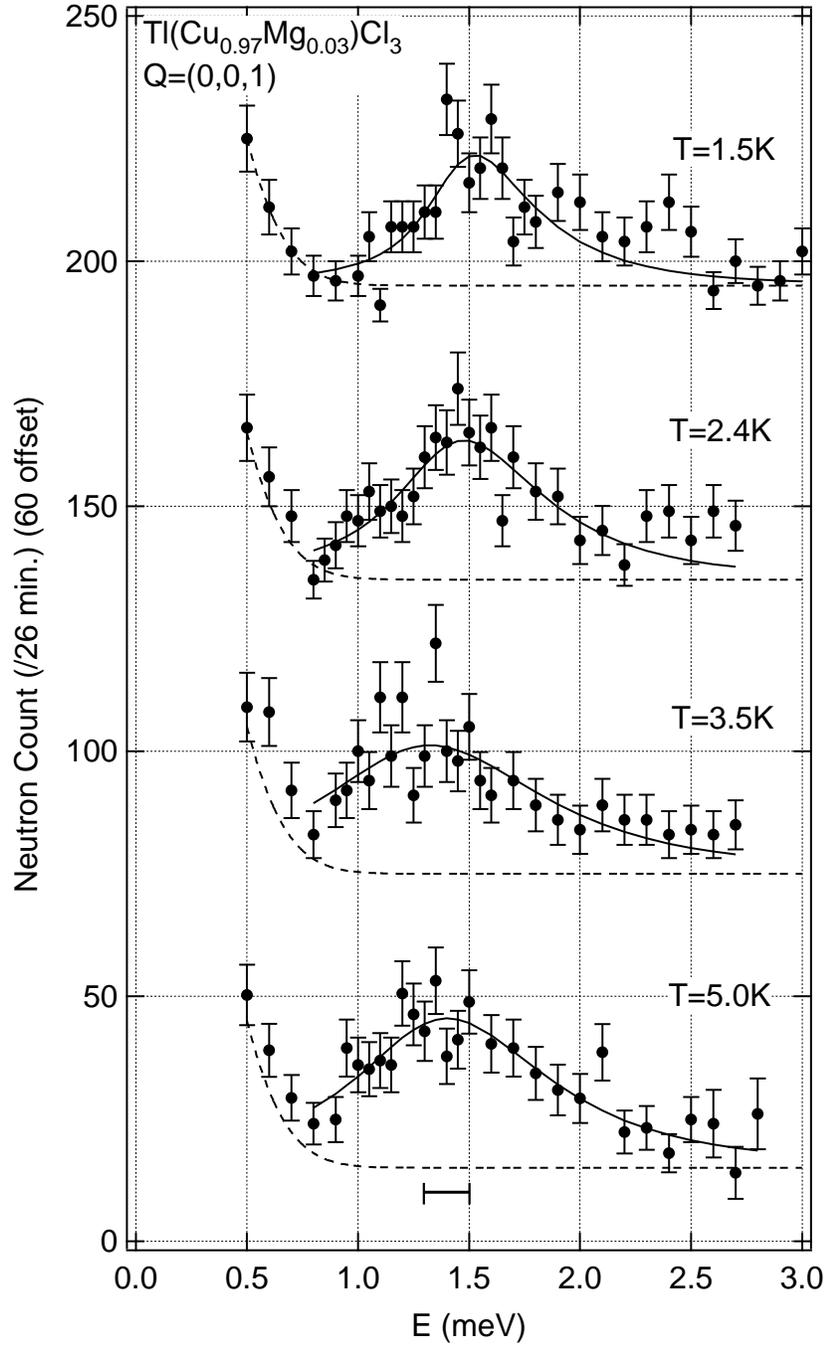}}
\vspace*{1cm}
	\caption{Constant-${\pmb Q}$ energy scan profiles in Tl(Cu$_{0.97}$Mg$_{0.03}$)Cl$_3$ at ${\pmb Q}=(0, 0, 1)$ for various temperatures. The solid lines are fits to the dispersion relations, as shown in Fig. \ref{ineladis}, convoluted with the instrumental resolution. The horizontal error bar indicates the calculated instrumental resolution widths. Dashed lines denote the background with the Bragg reflection at ${\pmb Q}=(0, 0, 1)$.}
	\label{001count}
\end{figure}

We have presented the results of neutron elastic and inelastic scattering on the doped spin gap system Tl(Cu$_{1-x}$Mg$_x$)Cl$_3$ with $x=0.03$. The magnetic Bragg reflections indicative of the impurity-induced antiferromagnetic ordering was observed at ${\pmb Q}=(h, 0, l)$ with integer $h$ and odd $l$ below $T_{\rm N}=3.45$ K. The spin structure in the impurity-induced ordered phase was determined, as shown in Fig. \ref{spinstructure}, with $\alpha=34.0^{\circ} \pm 4.7^{\circ}$. It was found that the spin structure of the impurity-induced antiferromagnetic ordered phase on average in Tl(Cu$_{0.97}$Mg$_{0.03}$)Cl$_3$ is the same as that of the field-induced magnetic ordered phase for ${\pmb H} \parallel b$ in the parent compound TlCuCl$_3$. The triplet magnetic excitation was clearly observed in the $a^*$-$c^*$ plane. The dispersion relations of the triplet excitation were determined, as shown in Fig. \ref{ineladis}. The main intradimer interaction and the effective interdimer interactions were obtained, as shown in Table \ref{table3}, using the RPA approximation. The lowest triplet excitation corresponding to the spin gap was also observed at ${\pmb Q}=(h, 0, l)$ with integer $h$ and odd $l$, as observed in TlCuCl$_3$. We found the rapid increase of the spin gap below $T_{\rm N}$ upon decreasing temperature. This indicates that the impurity-induced antiferromagnetic ordering coexists with the spin gap in Tl(Cu$_{0.97}$Mg$_{0.03}$)Cl$_3$. No antiferromagnetic spin-wave excitation could be observed.    

\acknowledgments

We would like to thank N. Metoki for experimental support on neutron scattering measurements and for stimulating discussions. The technical support of Y. Shimojo is also acknowledged. This work was supported by the Toray Science Foundation and a Grant-in-Aid for Scientific Research on Priority Areas (B) from the Ministry of Education, Culture, Sports, Science and Technology of Japan.

\begin{table}
\caption{Integrated Intensities of the magnetic Bragg reflections at $T=1.4$ K in Tl(Cu$_{0.97}$Mg$_{0.03}$)Cl$_3$. The intensities are normalized to the (0, 0, 1)$_{\rm M}$ reflection. $R$ is the reliability factor given by $R=\sum_{h,k,l}|I_{\rm cal}-I_{\rm obs}|/\sum_{h,k,l}I_{\rm obs}$.}
\label{table2}
\begin{center}
\begin{tabular}{@{\hspace{\tabcolsep}\extracolsep{\fill}}ccc}
\hline
$(h, k, l)$ & $I_{\rm obs}$ & $I_{\rm cal}$ \cr
\hline
(0, 0, 1)$_{\rm M}$ & 1 $\pm$ 0.021 & 1 \\
(0, 0, 3)$_{\rm M}$ & 0.006 $\pm$ 0.001 & 0.014\\
(0, 0, 5)$_{\rm M}$ & 0.058 $\pm$ 0.010 & 0.106\\
(1, 0, 1)$_{\rm M}$ & 0.039 $\pm$ 0.011 & 0.002\\
(1, 0, $-1$)$_{\rm M}$ & 0.059 $\pm$ 0.002 & 0.100\\
(1, 0, 3)$_{\rm M}$ & 0.024 $\pm$ 0.007 & 0.033\\
(1, 0, $-3$)$_{\rm M}$ & 0.352 $\pm$ 0.006 & 0.332\\
(2, 0, 1)$_{\rm M}$ & 0.017 $\pm$ 0.005 & 0.007\\
(2, 0, $-1$)$_{\rm M}$ & 0.065 $\pm$ 0.013 & 0.075\\
\hline
$R$ & 0.11\\
\hline
\end{tabular}
\end{center}
\end{table}

\begin{table}
\caption{The intradimer interaction $J$ and the effective interdimer interactions $J^{\rm eff}_{(lmn)}$ in TlCuCl$_3$ \cite{Oosawainela} and Tl(Cu$_{0.97}$Mg$_{0.03}$)Cl$_3$. All energies are in units of meV.}
\label{table3}
\begin{center}
\begin{tabular}{@{\hspace{\tabcolsep}\extracolsep{\fill}}ccc}
\hline
 & TlCuCl$_3$ \cite{Oosawainela} & Tl(Cu$_{0.97}$Mg$_{0.03}$)Cl$_3$ \\
\hline
$J$ & $5.68$ & $5.74$\\
$J^{\rm eff}_{(100)}$ & $-0.46$ & $-0.41$\\
$J^{\rm eff}_{(200)}$ & $0.05$ & $-0.03$\\
$J^{\rm eff}_{(1\frac{1}{2}\frac{1}{2})}$ & $0.49$ & $0.46$\\
$J^{\rm eff}_{(0\frac{1}{2}\frac{1}{2})}$ & $-0.06$ & $-0.03$\\
$J^{\rm eff}_{(201)}$ & $-1.53$ & $-1.41$\\
\hline
\end{tabular}
\end{center}
\end{table}

\newpage

\newpage

\newpage

\newpage

\newpage

\newpage

\newpage

\end{document}